\documentclass{elsart}
\usepackage{graphicx}
\usepackage{amssymb}\journal{Chemical Physics Letters}

\begin{document}

\begin{frontmatter}

\title{Toward the Stable Optical Trapping of a Droplet with Counter Laser Beams under Microgravity}

\author{Akihiro Isomura},
\author{Nobuyuki Magome\thanksref{nbunri}},
\author{Masahiro I. Kohira\thanksref{chudai}},
\author{Kenichi Yoshikawa\corauthref{cor}}
\corauth[cor]{Corresponding author. FAX: +81 75 753 3779}
\ead{yoshikaw@scphys.kyoto-u.ac.jp}

\thanks[nbunri]{Department of Food and Nutrition, Nagoya Bunri College,
 Nagoya 451-0077, Japan.}
\thanks[chudai]{Department of Physics, Chuo University, Kasuga,
 Bunkyo-ku, Tokyo 112-8551, Japan.}

\address{Department of Physics, Graduate School of Science, Kyoto
 University, Kyoto 606-8502, Japan \& Spatio-Temporal Project, ICORP, JST, JAPAN}

\begin{abstract}
To identify the optimum conditions for the optical trapping of a
droplet under microgravity, we theoretically analyzed the efficiency of
 trapping with counter laser beams. We found that the
distance between the two foci is an important parameter for obtaining
 stable trapping conditions. We also performed an optical trapping
experiment with counter laser beams under microgravity. The experimental results
correspond well to the theoretical prediction.
\end{abstract}

\end{frontmatter}

\section{Introduction}

Since the discovery of optical trapping in 1970 by
Ashkin\cite{ashkin_prl_1970}, optical tweezers have been actively applied in the
fields of biology, physical chemistry, condensed matter physics and so
on\cite{david_sci_2003}.
Recently, some reports have mentioned the advantages of such systems on the
International Space Station (ISS)\cite{panda_nasa_2002}\cite{susan_nasa_2004}.
The technique of optical trapping is expected to be useful on the
ISS for the manipulation of droplets on a ${\rm \mu m}$ to sub-${\rm mm}$
scale, including in crystal growth by avoiding the effect of
vessels. However, as far as we know, there has been no report on the
optical trapping of a droplet under microgravity conditions in air.

Recently, we reported the optical levitation of a droplet
under a linear increase in gravitational acceleration using a single laser\cite{kohira_cpl_2005}.
To attain stable optical trapping with a single beam, it is
necessary to use a lens with a high magnification. This means that
the working distance, or the distance between the lens surface and the
object, is rather small; on the order of a few ${\rm mm}$.
In view of such application on the ISS, it is
important to find the stable trapping condition with a greater working
distance. Toward this end, we adapted an optical
system with low-converged counter laser beams for optical trapping with
a long working distance.

\section{Theoretical}

We consider the counter laser system shown in Fig.1.
In this system, the distance between the foci $d$ plays a crucial
role. Although the qualitative property of such
counter laser systems has been discussed in Ref.\cite{rossen_optcomm_1977}, the effect of
the distance $d$ has not yet been discussed in detail. Here, we
calculate the trapping force produced by counter lasers
with the variables $d$ and $R$ (which is the radius of a spherical object).

\begin{figure}
\centering
\includegraphics[scale=0.40]{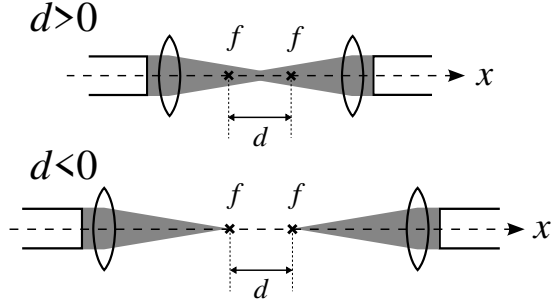}
\caption{Schematic illustration of the counter laser beams. $f$ is the focus of each laser beam and $d$ is
 the distance between them. We take the sign of $d$ as shown in the figure.}
\label{fig1}
\end{figure}

To take into account the effect of the inertia of the object, we
start our discussion with a motion equation.
To simplify the problem, we consider motion along the $x$-direction as in Fig.1.
\begin{equation}
m \ddot{x} = F_l + F_v + F_{others}
\end{equation}
where $m$ is the mass of the object, $\ddot{x}$ is the acceleration of
the object along the $x$-direction, $F_l$ is the force
along the optical axis induced by the converged laser, and $F_{others}$
is the other forces (including the effective
acceleration force along the $x$-direction and random force induced by air).

As we see in the latter part, the velocity of the trapped object $U$ is at most
${\rm 10^{-3} mm}$. We assume that the scale of the
trapped object $L$ is at most ${\rm 10^{-4} m}$, and the kinetic
viscosity of air $\nu$ is around ${\rm 10^{-5} m^2/s}$. In this condition,
the Reynolds number is: $Re = RU/\nu \simeq 10^{-2} \ll 1$ and we can
apply Stokes' approximation.
By calculating $m = 4\pi a^3 \rho /3$ and $F_{v} = 6 \pi \eta R v$, we can verify the relation $mv^2 / 2 \ll F_{v} \cdot \Delta x$; where the radius of the trapped
droplet $R$ is on the order of 10\ ${\rm \mu m}$, the experimental time resolution
$\Delta t$ is ${\rm 0.33\ s}$, the velocity of the trapped
droplet $v = \Delta x / \Delta t$, the viscosity of air $\eta = {\rm
1.8 \times 10^{-5}\ Pa \cdot s}$ and the density of the droplet $\rho= {\rm 1.0 \times
10^{3} kg/m^3}$.
This estimation indicates that the viscous force is large enough to
locate the droplet at an equilibrium position in the optical
potential. Therefore, under reasonable assumptions of the viscous
limit, we discuss the stability of trapping droplets through $F_l$ and the effective optical potential.

$F_l$ is found as follows. In the case of trapping a large object($R/
\lambda > 10$, where $\lambda$ is the wavelength of laser), we can
calculate the trapping force with the ray optics theory.
Within the framework of ray optics, a TEM$_{00}$
mode laser can be divided into rays, which are suffixed with $i$, and
each power of the ray $P_i$ is related to the beam deviation $\sigma$ and lens size $L$
 Each ray hits the surface of the droplet at a different incident angle $\phi_i$ ($0 \leq
\phi_i \leq \phi$), repeatedly reflects and is transmitted in the
droplet until the intensity of the rays decreases to the zero limit, and
the momentum is given with a certain efficiency along the $x$-axis,
$Q_i=Q_i(R,x;\phi_i,n_1,n_2)$:
\begin{eqnarray}Q_{i} = \sin \phi_i \left\{ R_i \sin 2 \theta_i -	\frac{T_i^2[\sin(2\theta_i - 2r_i)+R_i\sin2\theta_i]}{1+R_i^2+2R_i\cos2r_i}\right\}\nonumber\\+ \cos \phi_i \left\{ 1+R_i \cos 2\theta_i -\frac{T_i^2[\cos(2\theta_i-2r_i)+R_i\cos2\theta_i]}{1+R_i^2+2R_i\cos2r_i} \right\}
\label{eq:qi}
\end{eqnarray}
where $\theta_i$: incident angle, $r_i$: refractive angle,
$R_i$:reflection coefficient and $T_i$: transmission coefficient. These
parameters are obtained by considering the geometric relation between
the droplet and the direction of the beam (see
ref.\cite{ashkin_bpj_1992}). Assuming a spherical droplet with radius
$R$ whose center is located at distance $x$ from the center of the
foci, we calculate the details of the reflections and transmissions for
all paths of the laser beam.
The total force $f_l$ is 
\begin{equation}
f_l = \sum_{i} \frac{n_1 P_i}{{\rm c}}Q_i = \frac{n_1 P}{{\rm c}} Q_x
\end{equation}
where $Q_x=Q_x(R,x; \phi,\frac{\sigma}{L},n_1,n_2)$: trapping efficiency
along the optical axis and $c$: velocity of light. In addition,
considering the distances between the foci, the total force $F_l$ is:
\begin{equation}
F_l(d,x,R) =f_{l}^{Right} + f_{l}^{Left}
\end{equation}
Under the assumption of a spherical trapped object, optical trapping exhibits
geometrical symmetry and we can use a normalized unit of length.
Eventually, the optical potential is expressed by two normalized parameters, $d/R$
and $x/R$.

\begin{figure}
\centering
\includegraphics[scale=0.80]{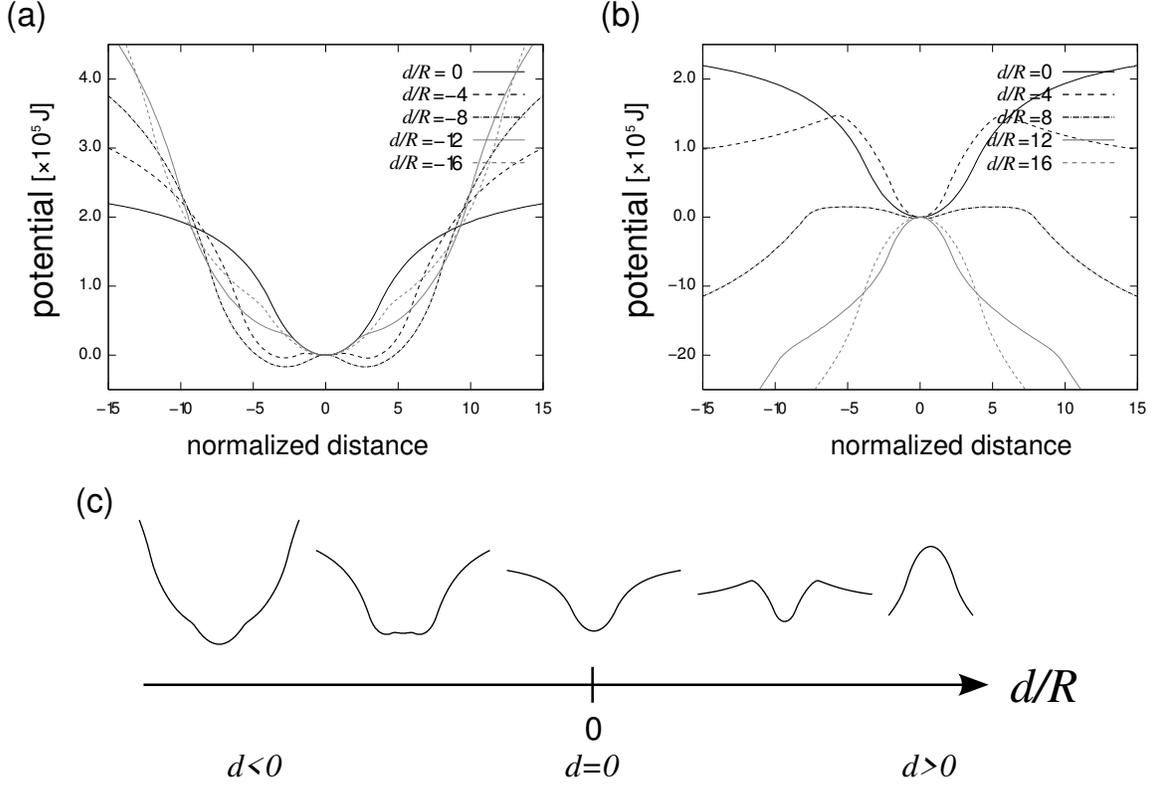}
\caption{Calculated potential of the counter beam lasers, deduced by
 a theoretical calculation. (a) $d<0$. (b) $d>0$. In the
 theoretical calculation, the following parameters were used: laser
 power $P {\rm = 150\ mW}$, convergence angle $\phi$ = 17.0 degrees,
 injected ratio of laser to the lens $\sigma/L = 1.5$, refractive index
 of the medium $n_1 = 1.00$,
 refractive index of the droplet $n_2 = 1.35$, density of the droplet
 $\rho = 9.65 \times {\rm 10^2\ kg/m^3}$.
 (c)Schematic illustration of the calculated
 potential ((a) and (b)) made by the two-beam lasers, considering the radius
 of the droplet and the disagreement between the axes. }
\end{figure}

The results of the calculation for effective trapping potential are shown
in Fig.2, indicating that the trapping efficiency is significantly dependent on
$d$. An optical configuration with positive $d$ is clearly preferable for optilcal trapping under microgravity.
Figure 2(c) shows the change in the potential profile as a
function of $d/R$. As shown, not only the distance between
the foci but also the radius of the droplet greatly affect the profile of the
optical potential. This chart is reasonable for understanding qualitative
properties: if the radius $R$ is sufficiently large, the trapped object is insensitive to
the distance between the foci $d$ and the sign of $d$ become less
important. In contrast, when $R$ is sufficiently small, $d$ becomes
non-negligible and we must choose better settings of $d$. Since we
sought to define the optimum conditions for trapping large objects (sub-mm scale), this diagram is
applicable for designing a counter laser trapping system. For example. if we can precisely adjust the positions of the
foci, the setting at $d/R = 4$ is advantageous for trapping the object
rather than $d/R = -4$. The potentiality to trap small particles is important
in the growth of protein
crystals\cite{giege_pcgcm_1995}\cite{hosokawa_jap_2004} and liquid
droplets\cite{magome_jpc_2003}.

\section{Experimental}

Microgravity conditions were achieved by using a jet
airplane(Mitsubishi MU-300, operated by Diamond Air Service Co., Aichi,
Japan). In the experiment, the airplane flew along a parabolic
flight profile, as shown in Fig.3. During parabolic flight, the
effective gravitational acceleration $g$ in the airplane was 0.01 $g_0$
for about 20 s, where the gravitational acceleration on the
ground $g_0 = 9.8 {\rm m/s^2}$.
\begin{figure}
\centering
\includegraphics[scale=0.80]{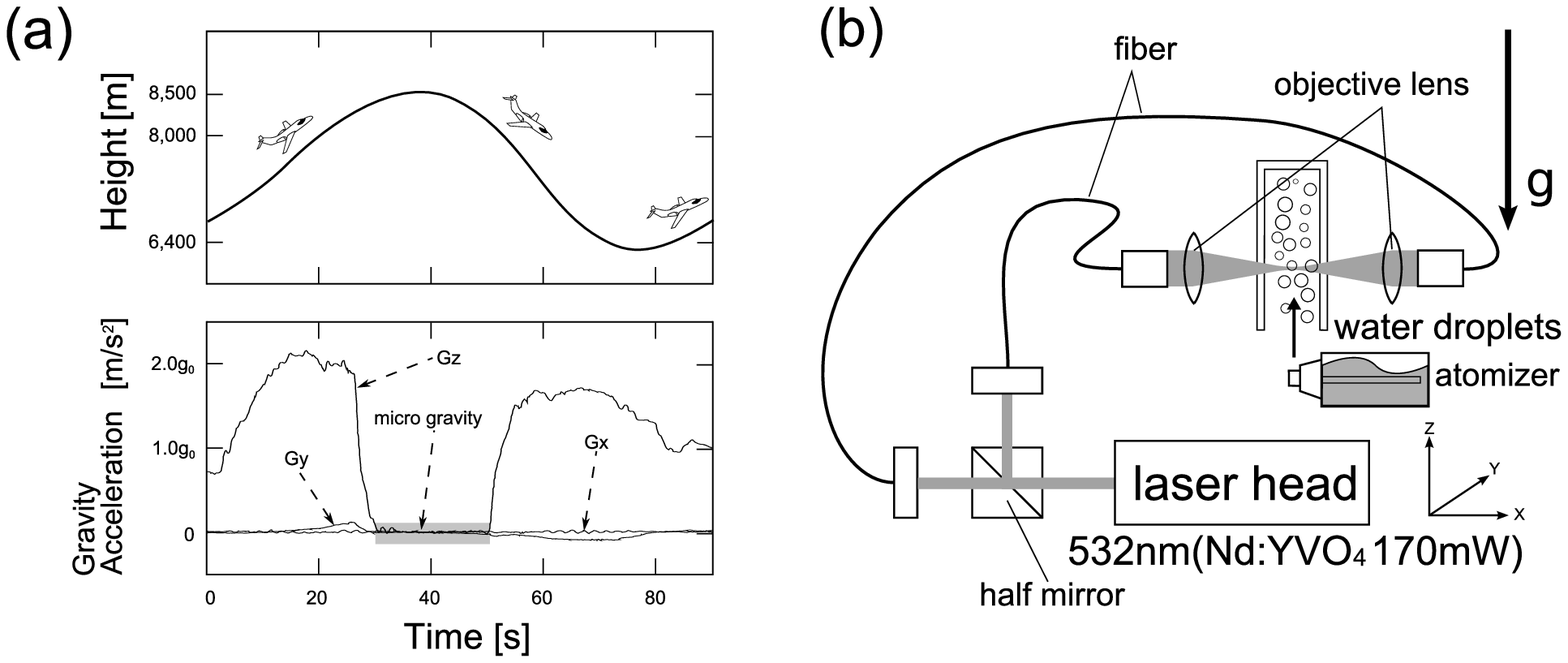}
\caption{(a) Schematic illustration of parabolic flight (top) and the
 actual change in gravity with the present parabolic
 flight profile(bottom). (b) Schematic illustration of the optical trapping
 system.}
\end{figure}

The experimental setup for the optical trapping is shown
in Fig.3(b). We selected a counter laser system to press the droplets to
the center of the foci.\cite{rossen_pl_1976}\cite{cizmar_apl_2005} The laser beam from a diode pumped
${\rm Nd:YVO_4}$($\lambda = 532\ {\rm nm}$) was split into two beams by a half
mirror, transferred via optical fibers, and focused by achromatic lenses. The
convergence angle of the lenses was 17 degrees, so the working distance
was on the order of {\rm cm}. The laser power was set to {\rm 170 mW}.
We arranged two different optical settings, where the positions of
the laser foci are different, either $d$ is positive or negative as shown in Fig.1.
The droplets were injected into a glass cell ${\rm (10 mm \times 10 mm \times 50 mm)}$
with an atomizer. The droplets, as visualized by scattering visible laser light, were monitored with a CCD camera
from the y-direction in Fig.3(b). Movies were recorded at 30 frames per second. The experiments was carried out at arround 298K.

\section{Results and Discussion}

On the ground in air, using the same optical trapping system, the
injected droplets immediately fell due to the force of gravity. However,
under microgravity in air, most of the droplets didn't
fall and we could observe the motions of the droplets along the laser axes.
To trace the droplets' motion  along the laser axes, we converted
the movie images to a spatio-temporal diagram. 
Figure 4 shows the results of the optical trapping of the droplets under
microgravity.

\begin{figure}
\centering
\begin{tabular}{c}
\vspace{0.75cm}
\includegraphics[scale=0.9]{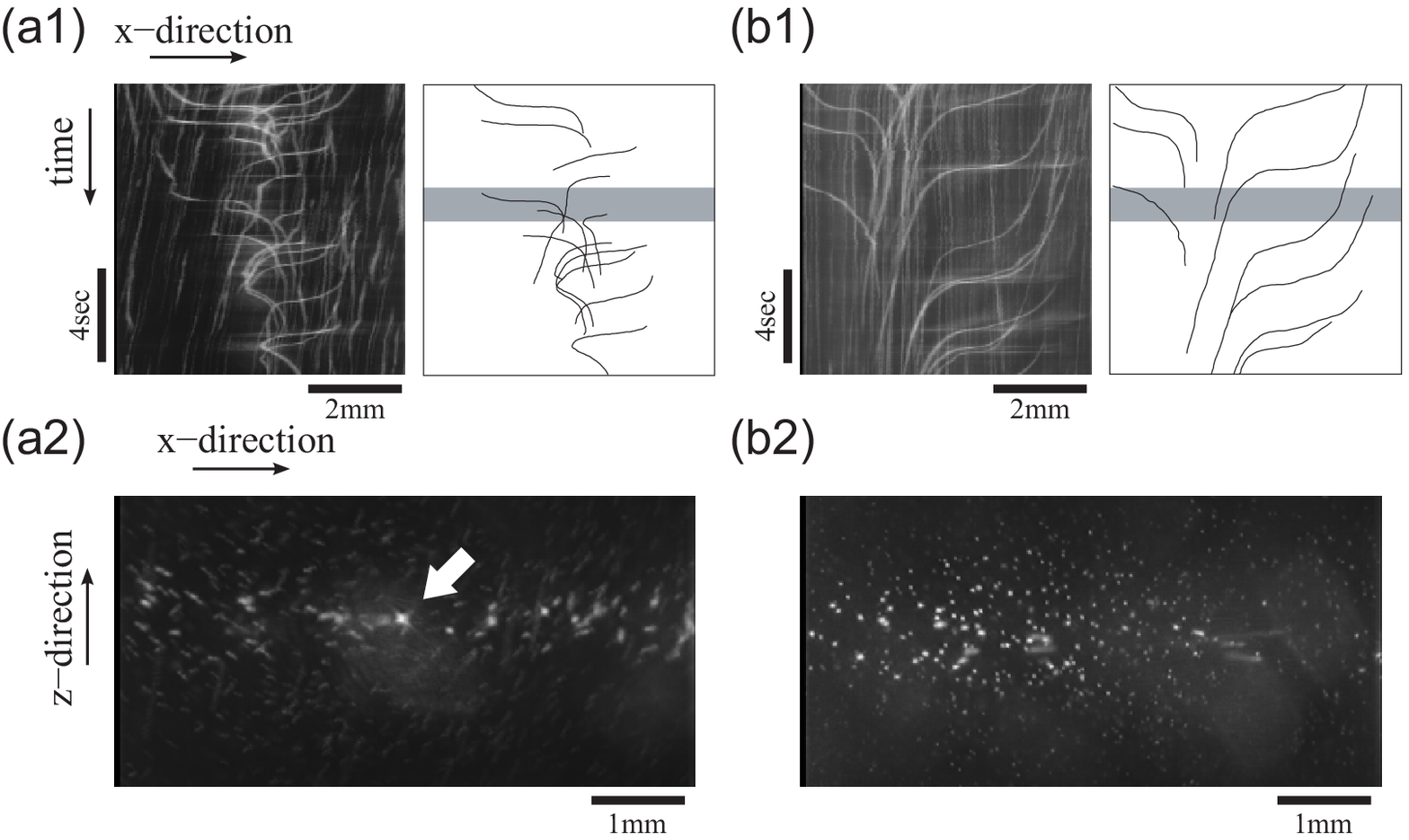}\\
\includegraphics[scale=0.8]{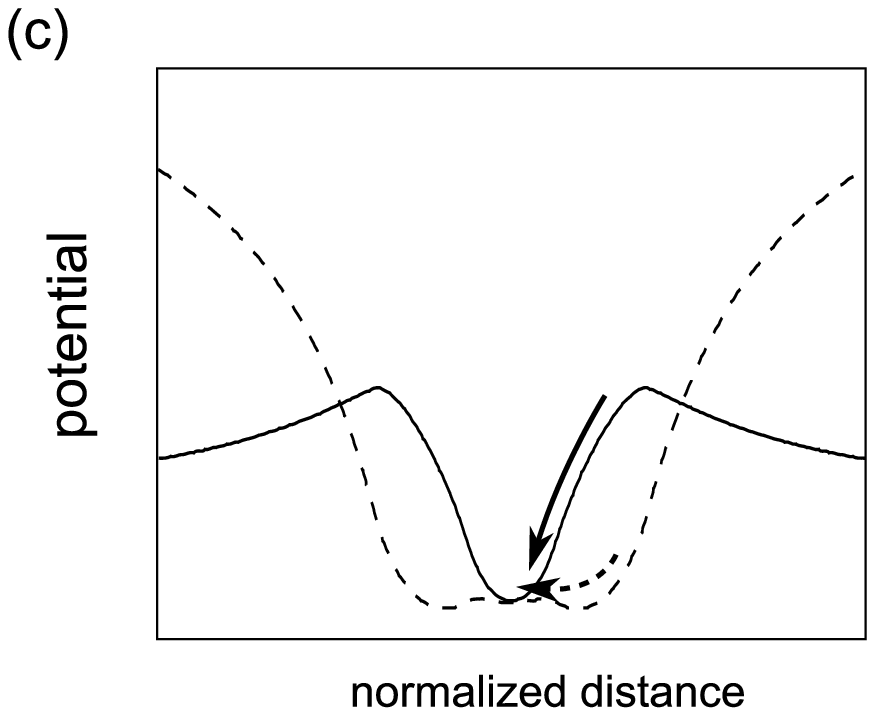}\\
\end{tabular}
\caption{Spatio-temporal diagrams of droplets under microgravity conditions
 together with selected representations of traces. The optics are
 (a1) $d>0$. (b1) $d<0$. 
(a2) and (b2) are the 1 sec accumulation of the video frames, indicating
 the existence of a trapped particle in the former, whereas no particle
 was fixed in the latter.
 (c) Schematic illustration to explain the
 difference between trapped and untrapped droplets. The solid line is
 the potential profile in the case of $d/R = 4$ and the broken line is
 that in the case of $d/R = -4$. }
\end{figure}
Figure 4 (a1) and (b1) are spatio-temporal diagrams which show the
trajectory of the droplets. 
Figure 4 (a2) and (b2) are snapshots of the droplets. 
In these figures, the burring on the images correspond to rather fast motion of the droplets.
The small white spots are droplets that have departed from the laser
axes and are floating in front of or behind the camera foci.
In contrast, a trapped droplet is a star-like object(white arrow), since
a trapped droplet on the laser axes scatter light intensively.

Figure 4(c) shows typical profiles depicted based on the results of Fig.2(a) and (b),
which are associated with the two different optical
settings in Fig.4(a2) and (b2). As shown near the trapping
region, the gradient force in the case of the solid arrow is much greater
than that in the case of the dashed arrow($d/R=-4$). It is clear that
our theoretical study reproduces the experimental trends well.

\section{Conclusion}
We performed the optical trapping of a water droplet with a counter laser
in air under microgravity and investigated the properties of the counter
laser trapping system. We showed that the position of the laser foci strongly affected the trapping
efficiency. A theoretical calculation based on ray optics reproduced the
experimental trend. The present results may contribute to the design of
a manipulation system on the International Space Station (ISS),
including experiments on protein crystal growth and a container-less micro-reactor.

\section{Acknowledgement}
The authors thank Mr. S. Watanabe, Mr. Y. Sumino, and N. Matsuda for
their helpful suggestions and
Ms. Hayata and Messrs. Fujii, Kawakatsu, and Takahashi for their technical
assistance. This research was supported in part by a Grant-in Aid from
the Ministry of Education, Science, Sports and Culture of Japan, the
Grant-in-Aid for the 21st Century COE ¡ÈCenter for Diversity and
Universality in Physics¡É from the Ministry of Education, Culture,
Sports, Science and Technology (MEXT) of Japan, and ¡ÈGround-based
Research Announcement for Space Utilization¡É promoted by the Japan
Space Forum.

\bibliographystyle{elsart-num}
\bibliography{microg}

\begin{thebibliography}{10}
\expandafter\ifx\csname url\endcsname\relax
  \def\url#1{\texttt{#1}}\fi
\expandafter\ifx\csname urlprefix\endcsname\relax\def\urlprefix{URL }\fi

\bibitem{ashkin_prl_1970}
A.~Ashkin, Acceleration and trapping of particles by radiation pressure, Phys.
  Rev. Lett. 24 (1970) 156.

\bibitem{david_sci_2003}
D.~G. Grier, A revolution in optical manipulation, Nature 424 (2003) 810.

\bibitem{panda_nasa_2002}
J.~Panda, C.~R. Gomez, Setting up a rayleigh scattering based flow measuring
  system in a large nozzle testing facility, NASA/TM (2002) 211985.

\bibitem{susan_nasa_2004}
S.~Y. Wrbanek, K.~E. Weiland, Optical levitation of micro-scale particles in
  air, NASA/TM (2004) 212889.

\bibitem{kohira_cpl_2005}
M.~I. Kohira, A.~Isomura, N.~Magome, S.~Mukai, K.~Yoshikawa, Optical levitation
  of a droplet under a linear increase in gravitational acceleration, Chem.\
  Phys. \ Lett. 414 (2005) 389.

\bibitem{rossen_optcomm_1977}
G.~Roosen, A theoretical and experimental study of the stable equilibrium
  positions of spheres levitated by two horizontal laser beams, Opt.\ Comm. 21
  (1977) 189.

\bibitem{ashkin_bpj_1992}
A.~Ashkin, Forces of a single-beam gradient laser trap on a dielectric sphere
  in the ray optics regime, Biophys.\ J. 61 (1992) 569.

\bibitem{giege_pcgcm_1995}
R.~Gieg\'{e}, J.~Drenth, A.~Ducruix, A.~McPherson, W.~Saenger, Study of the
  aerodynamic trap for containerless laser materials processing in
  microgravity, Prog.\ Cryst.\ Growth\ Charact.\ Mater. 30 (1995) 237.

\bibitem{hosokawa_jap_2004}
Y.~Hosokawa, S.~Matsumura, H.~Masuhara, K.~Ikeda, A.~Shimo-oka, H.~Mori, Laser
  trapping and patterning of protein microcrystals: Toward highly integrated
  protein microarrays, J.\ Appl.\ Phys. 96 (2004) 2945.

\bibitem{magome_jpc_2003}
N.~Magome, M.~I. Kohira, E.~Hayata, S.~Mukai, K.~Yoshikawa, Optical trapping of
  a growing water droplet in air, J. \ Phys. \ Chem. \ B 107 (2003) 3988.

\bibitem{rossen_pl_1976}
G.~Roosen, C.~Imbert, Optical levitation by means of two horizontal laser
  beams: A theoretical and experimental study, Phys.\ Lett. 59A (1976) 6.

\bibitem{cizmar_apl_2005}
T.~Cizmar, V.~Garc\'{e}s-Ch\'{a}vez, K.~Dholakia, P.~Zem\'{a}nek, Optical
  conveyor belt for ddelivery of submicron objects, Appl.\ Phys.\ Lett. 86
  (2005) 17401.

\end{thebibliography}
\end{document}